\documentclass[pre,preprint,superscriptaddress,amsmath,amssymb,floatfix]
{revtex4}
\usepackage{graphics}
\usepackage[dvips]{graphicx}
\usepackage{amsfonts}
\usepackage{eufrak}
\usepackage{graphicx}
\usepackage[mathscr]{eucal}
\begin{document}
\title{Gas-liquid critical point in ionic fluids}
 \author{O. Patsahan }
\address{Institute for Condensed Matter Physics of the National
Academy of Sciences of Ukraine, 1 Svientsitskii Str., 79011 Lviv, Ukraine}
 \author{I. Mryglod}
\address{Institute for Condensed Matter Physics of the National
Academy of Sciences of Ukraine, 1 Svientsitskii Str., 79011 Lviv, Ukraine}
\author{T. Patsahan}
\address{Institute for Condensed Matter Physics of the National
Academy of Sciences of Ukraine, 1 Svientsitskii Str., 79011 Lviv, Ukraine}
 \date{\today}

\begin{abstract}
Based on the method of collective variables we develop  the statistical field theory for the study of a simple charge-asymmetric $1:z$ primitive model (SPM).  It is shown that the well-known approximations for the free energy, in particular DHLL and  ORPA, can be obtained within the framework of this theory. In order to study the gas-liquid critical point of SPM we propose the method  for the calculation of  chemical potential conjugate to the total number density which   allows us to take into account the higher order fluctuation effects. As a result, the gas-liquid phase diagrams are calculated for $z=2-4$. The results demonstrate the  qualitative agreement  with MC simulation data: critical temperature decreases when $z$ increases and critical density increases rapidly with $z$.

\end{abstract}

\maketitle

\section{Introduction}
Besides being of  fundamental interest, ionic systems including electrolyte solutions, molten salts and ionic liquids  deserve great attention from practical point of view. For example, new ionic liquids with  very low vapor pressure may find applications in environmentally clean industry. Over the last ten years the phase and critical behavior of ionic fluids has been a subject of intense research.   For reviews of  experimental and theoretical situation see Refs. \cite{levelt1,fisher1,fisher2,stell1,stell2,ciach:00:0,Schroer:review} and the references cited therein.   Basic properties of ionic systems can be described by the primitive model (PM) consisting of a mixture of $m$ species of charged hard spheres. A two-component mixture of positive $q_{0}$ and negative $-zq_{0}$ charges  having all the same diameter $\sigma$ is a simple charge-asymmetric PM (SPM).  In the case $z=1$ we have a well-known restricted primitive model (RPM). Early studies \cite{stellwularsen} established that  the RPM has a gas-liquid (GL)-like critical point. A reasonable theoretical description of the GL critical point of the RPM was accomplished at
a mean-field (MF) level using integral equation methods \cite{stell1,stell3} and Debye-H\"{u}ckel (DH)
theory \cite{levinfisher}. Over the last decade numerous simulation studies have been devoted to the location of GL phase transition  of this model
\cite{panagiotopoulos3,caillol2,panagiotopoulos2,caillol1,orkoulas1,ydp,panagiotopoulos1,caillol_mc,luijten}
and the most reliable current estimations turn out to be near $T_{c}^{*}=0.049$, $\rho_{c}^{*}=0.08$ when the temperature $T^{*}$ and the density $\rho^{*}$ are in standard dimensionless units.
 Due to controversial experimental findings, the critical behavior of the RPM has also been under active debates  \cite{fisher3,schroer,valleau,camp,luijten1} and  strong evidence  for an Ising universal class has been found by recent  simulations \cite{luijten,kim,kim04}  and theoretical  \cite{ciach:00:0, ciach:05:0,ciach:06:1,patsahan:04:1} studies.

 In spite of significant progress in this field,  ionic systems are far from being completely understood. The investigation of more complex models is very important in understanding the nature of critical and phase behavior of real ionic fluids which demonstrate both the charge and size asymmetry as well as other complexities such as  short-range attraction.
The studies of the effects of  charge-asymmetry on the phase diagram have  been recently started using  both the computer simulations \cite{yan-pablo:02, panagiot-fisher:02,caillol_mc} and theoretical methods \cite{netz_orland,caillol_1,aqua_banerjee_fisher,fisher_aqua_banerjee}. Monte Carlo (MC) simulations show that SPM exhibits a GL-like phase transition, specifically $T_{c}^{*}(z)$  decreases with the increase of  $z$ while $\rho_{c}^{*}(z)$ rapidly increases. Unfortunately, the results obtained within the framework of both the original DH theory and the mean spherical approximation (MSA) are independent of $z$. The  results for the $z$-dependence of the critical temperature, obtained within the framework of the field-theoretical analysis  \cite{netz_orland,caillol_1}  contradict the simulations displaying the increase of temperature with $z$. On the other hand,  the  recent results found based on  the DH theory incorporated Bjerrrum ion pairs and their solvation in the residual ionic fluid  \cite{aqua_banerjee_fisher,fisher_aqua_banerjee} (theories DHBjCI and DHBjCIHC) demonstrate the decrease of $T_{c}^{*}$ with charge asymmetry for $z=1-3$ which  agrees with MC data. This implies "... that recognizing ionic association is inescapable for a successful theory"  \cite{fisher_aqua_banerjee}.

The motivation of this paper is to check the above statement by means of the theory that exploites the method of collective variables (CVs) \cite{yukhol}. Our aim is to answer the question: is the theory, which does not include  the effects of association explicitely in the Hamiltonian,  capable of correctly describing, at least on the qualitative level, the GL-like phase diagram of SPM?
To this end,  we  develop a theoretical approach which is based on the functional representation of  configurational Boltzmann factor in terms of  CVs.
 For a general case of a two-component model which includes both the short- and long-range interactions we derive an exact expression for the functional integral of the grand partition function (GPF). The  CV action obtained depends upon the two sets of CVs: $\left\lbrace \rho_{{\mathbf k}}\right\rbrace $ (and  conjugate $\left\lbrace \omega_{{\mathbf k}}\right\rbrace $) and $\left\lbrace Q_{{\mathbf k}}\right\rbrace $ (and  conjugate $\left\lbrace \gamma_{{\mathbf k}}\right\rbrace $) describing the total number density and charge density fluctuations, respectively. We start with  the saddle point (or MF) solutions $\bar\rho$ (and $\bar\omega$) and $\bar Q$ (and $\bar\gamma$),  and then we expand  the CV action functionally around the MF solutions. As a result, we obtain an infinite expansion in terms
 of CVs $\delta\rho_{{\mathbf k}}$ and $\delta Q_{{\mathbf k}}$ (or $\delta\omega_{{\mathbf k}}$ and $\delta \gamma_{{\mathbf k}}$).  For SPM, based on the GPF in the Gaussian approximation, we propose  a method for the calculation of  chemical potential conjugate to the total number density.   The method  allows us to take into account the higher order fluctuation effects and consists in solving the equations for chemical potentials by means of successive approximations that correspond to the contributions of the higher-order correlations. Its initial idea  to some extent resembles the idea sketched  out by Hubbard in \cite{hubbard}.

 It should be noted that  the CVs were  first introduced  in the 1950s \cite{bohm,zubar} and then on their basis the so-called CVs method was developed \cite{jukh,yukhol}. Nearly at that time, other functional approaches based on the Stratonovich-Hubbard transformation \cite{stratonovich,hubbard} were originated. As was shown recently \cite{caillol_patsahan_mryglod} both groups of theories  are  in  close relation.

The layout of the paper is as follows. In Section~2, starting from the Hamiltonian of an asymmetric two-component model with  long- and short-range interactions, we derive the exact functional representation of the GPF. We obtain an explicit expression for free energy of SPM in the random phase  approximation (RPA) that  leads to the well-known results, i.e. the DH limiting law (DHLL)  for the point charge particles  and the free energy in the optimized random phase approximation (ORPA) (or MSA) for SPM.  In Section~3 we study  the GL critical point  of  SPM taking into account the correlation effects of higher order.  The results obtained demonstrate qualitative agreement with the MC simulation data for both the critical temperature and the critical density. We conclude in Section~4.

\section{Functional representation of the grand partition function of a two-component ionic model in the  method of CVs}
Let us consider a general case of a classical two-component system consisting of $N$ particles among which there exist $N_{1}$ particles of species $1$ and $N_{2}$ particles of species $2$. The pair interaction potential is assumed to be of the following form:
\begin{equation}
U_{\alpha\beta}(r)=\phi_{\alpha\beta}^{HS}(r)+\phi_{\alpha\beta}^{C}(r)+\phi_{\alpha\beta}^{SR}(r),
\label{2.1a}
\end{equation}
where $\phi_{\alpha\beta}^{HS}(r)$ is the interaction potential between the two  additive hard spheres of diameters $\sigma_{\alpha\alpha}$ and $\sigma_{\beta\beta}$. We call the two-component hard sphere system a reference system (RS). Thermodynamic
and structural properties of RS are assumed to be known.
 $\phi_{\alpha\beta}^{C}(r)$ is the Coulomb potential: $\phi_{\alpha\beta}^{C}(r)=q_{\alpha}q_{\beta}\phi^{C}(r)$, where $\phi^{C}(r)=1/(D r)$, $D$ is the dielectric constant,  hereafter we put $D=1$. The solution is made of both positive and negative ions so that the electroneutrality is satisfied,
$\sum_{\alpha=1}^{2}q_{\alpha}c_{\alpha}=0$,
and $c_{\alpha}$ is the concentration of the species $\alpha$, $c_{\alpha}=N_{\alpha}/N$.
The ions of the species $\alpha=1$ are characterized by their hard sphere diameter $\sigma_{11}$ and their electrostatic charge $+q_{0}$ and those of species $\alpha=2$, characterized by diameter $\sigma_{22}$, bear opposite charge $-zq_{0}$ ($q_{0}$ is elementary charge and $z$ is the parameter of charge asymmetry).
  $\phi_{\alpha\beta}^{SR}(r)$ is the potential of the short-range interaction:
$\phi_{\alpha\beta}^{SR}(r)=\phi_{\alpha\beta}^{R}(r)+\phi_{\alpha\beta}^{A}(r)$, where $\phi_{\alpha\beta}^{R}(r)$ is used to mimic the soft core asymmetric repulsive interaction, $\phi_{\alpha\beta}^{R}(r)$ is assumed to have a Fourier transform; $\phi_{\alpha\beta}^{A}(r)$ describes a van der Waals-like attraction.

We consider the grand partition function (GPF) of the system which can be written as follows:
\begin{equation}
\Xi[\nu_{\alpha}]=\sum_{N_{1}\geq 0}\sum_{N_{2}\geq
0}\prod_{\alpha=1,2}
\frac{\exp(\nu_{\alpha}N_{\alpha})}{N_{\alpha}!} \int({\rm
d}\Gamma) \exp\left[-\frac{\beta}{2}\sum_{\alpha\beta}\sum_{ij}
U_{\alpha\beta}(r_{ij})\right].
\label{2.1}
\end{equation}
Here the following notations are used:
$\nu_{\alpha}$ is the dimensionless chemical potential, $\nu_{\alpha}=\beta\mu_{\alpha}-3\ln\Lambda$, $\mu_{\alpha}$ is the chemical potential of the $\alpha$th species, $\beta$ is the reciprocal temperature,
$\Lambda^{-1}=(2\pi m_{\alpha}\beta^{-1}/h^{2})^{1/2}$ is the inverse de Broglie thermal wavelength; $(\rm d\Gamma)$ is the element of configurational space of the particles.

Let us introduce operators $\hat\rho_{{\mathbf k}}$ and $\hat Q_{{\mathbf k}}$
\[
\hat\rho_{{\mathbf k}}=\sum_{\alpha}\hat\rho_{{\mathbf k},\alpha}
\qquad \hat Q_{{\mathbf
k}}=\sum_{\alpha}q_{\alpha}\hat\rho_{{\mathbf k},\alpha},
\]
which are  combinations of the Fourier transforms of the microscopic number density of the species  $\alpha$:  $\hat\rho_{{\mathbf k},\alpha}=\sum_{i}\exp(-{\rm i}{\mathbf
k}{\mathbf r}_{i}^{\alpha})$. In this case  a part of the Boltzmann factor in (\ref{2.1}) which does not include the RS interaction can be presented as follows:
\begin{eqnarray}
&&\exp\left[-\frac{\beta}{2}\sum_{\alpha\beta}\sum_{i,j}(U_{\alpha\beta}(r_{ij})-\phi_{\alpha\beta}^{HS}
(r_{ij}))\right]=\exp\left[-\frac{1}{2}\sum_{{\bf k}}(\tilde\Phi_{NN}\hat\rho_{{\mathbf k}}\hat\rho_{{\mathbf -k}}\right.  \nonumber \\
&&
\left.+\tilde\Phi_{QQ}\hat Q_{{\mathbf k}}\hat Q_{{\mathbf -k}}+ 2\tilde\Phi_{NQ}\hat\rho_{{\mathbf k}}\hat Q_{{\mathbf -k}})+\frac{1}{2}\sum_{\alpha}N_{\alpha}\sum_{{\mathbf
k}}(\tilde\Phi_{\alpha\alpha}^{SR}(k)+q_{\alpha}^{2}\tilde\Phi^{C}(k))\right],
\label{2.2}
\end{eqnarray}
where
\begin{eqnarray}
\tilde{\Phi}_{NN}(k)&=&\frac{1}{(1+z)^{2}}\left[
z^{2}\tilde{\Phi}_{11}^{SR}(k) + 2z \tilde{\Phi}_{12}^{SR}(k)
+\tilde{\Phi}_{22}^{SR}(k)\right]  \nonumber \\
\tilde{\Phi}_{QQ}(k)&=&\frac{1}{(1+z)^{2}}\left[
\tilde{\Phi}_{11}^{SR}(k)
-2\tilde{\Phi}_{12}^{SR}(k)+\tilde{\Phi}_{22}^{SR}(k)\right]
+\tilde{\Phi}^{C}(k) \nonumber \\
\tilde{\Phi}_{NQ}(k)&=&\frac{1}{(1+z)^{2}}\left[ z
\tilde{\Phi}_{11}^{SR}(k)
+(1-z)\tilde{\Phi}_{12}^{SR}(k)-\tilde{\Phi}_{22}^{SR}(k)\right]
\label{2.3}
\end{eqnarray}
and we use the notations $\tilde{\Phi}_{\alpha\beta}^{X\ldots}(k)=\frac{\beta}{V}\tilde\phi_{\alpha\beta}^{X\ldots}(k)$ with
$\tilde{\phi}_{\alpha\beta}^{X\ldots}(k)$ being a Fourier transform of the corresponding interaction potential.

In order to introduce the collective variables (CVs) we use  the identity
\begin{equation}
\exp\left[-\frac{1}{2}\sum_{{\bf k}}\tilde\Phi\hat\eta_{{\mathbf k}}\hat\eta_{-{\mathbf k}}\right] =\int({\rm
d}\eta)\delta_{{\cal
F}}[\eta-\hat\eta]\exp\left[-\frac{1}{2}\sum_{{\bf k}}\tilde\Phi\eta_{{\mathbf k}}\eta_{-{\mathbf k}}\right],
\label{2.4}
\end{equation}
where $\delta_{{\cal F}}[\eta-\hat\eta]$ denotes the functional delta function
\[
\delta_{{\cal F}}[\eta-\hat\eta]\equiv\int({\rm d}\lambda)\exp\left[{\rm i}\sum_{{\bf k}}\lambda_{{\mathbf k}}(\eta-\hat\eta_{{\mathbf k}})\right],
\]
 $\eta_{{\mathbf k}}=\eta_{{\mathbf k}}^c-{\rm i}\eta_{{\mathbf k}}^s$ ($\eta_{{\mathbf k}}=\rho_{{\mathbf k}},Q_{{\mathbf k}}$) is the collective variable and
\[
({\rm d}\eta)={\rm d}\eta_{0}\prod_{{\mathbf k\neq 0}}{\rm d}\eta_{{\mathbf k}}^{c}{\rm d}\eta_{{\mathbf k}}^{s}, \qquad
({\rm d}\lambda)={\rm d}\lambda_{0}\prod_{{\mathbf k\neq 0}}{\rm d}\lambda_{{\mathbf k}}^{c}{\rm d}\lambda_{{\mathbf k}}^{s}.
\]
The indices $c$ and $s$ denote real and imaginary parts of $\eta_{{\mathbf k}}$ ($\lambda_{{\mathbf k}}$), respectively, the product over ${\mathbf k}$ is performed in the upper semi-space.

Taking into account (\ref{2.2})-(\ref{2.4}),  we can rewrite (\ref{2.1})
\begin{equation}
\Xi[\nu_{\alpha}]=\int ({\rm d}\rho)({\rm d}Q)({\rm
d}\omega)({\rm d}\gamma) \exp\left(-{\cal H}[\nu_{\alpha},\rho,Q,\omega,\gamma] \right),
\label{2.5}
\end{equation}
where the CV action ${\cal H}$ is as follows:
\begin{eqnarray}
&&{\cal H}[\nu_{\alpha},\rho,Q,\omega,\gamma]=\frac{1}{2}\sum_{{\mathbf
k}}[\tilde \Phi_{NN}(k)\rho_{{\mathbf k}}\rho_{-{\mathbf
k}}+\tilde \Phi_{QQ}(k)Q_{{\mathbf k}}Q_{-{\mathbf k}}+2\tilde \Phi_{NQ}(k) \nonumber\\
&&
\times\rho_{{\mathbf k}}Q_{-{\mathbf
k}}]
-{\rm i}\sum_{{\mathbf k}}(\omega_{{\mathbf k}}\rho_{{\mathbf
k}}+\gamma_{{\mathbf k}}Q_{{\mathbf k}})-\ln \Xi_{HS}[\bar
\nu_{\alpha};-{\rm i}\omega,-{\rm i}q_{\alpha}\gamma].
\label{2.5a}
\end{eqnarray}
In (\ref{2.5a}) CVs $\rho_{{\mathbf k}}$ and $Q_{{\mathbf k}}$ describe fluctuations of the total number density and charge density, respectively.
$\Xi_{HS}[\bar\nu_{\alpha};-{\rm i}\omega,-{\rm i}q_{\alpha}\gamma]$ is the GPF of a two-component system of  hard spheres with the renormalized chemical potential $\bar \nu_{\alpha}$ in the presence of the local field $\psi_{\alpha}(r_{i})$
\begin{eqnarray}
\Xi_{HS}[\ldots]&=&\sum_{N_{1}\geq 0}\sum_{N_{2}\geq 0}\prod_{\alpha=1,2}
\frac{\exp(\bar\nu_{\alpha}N_{\alpha})}{N_{\alpha}!}
\int({\rm d}\Gamma)
\exp\left[-\frac{\beta}{2}\sum_{\alpha\beta}\sum_{ij}
\phi_{\alpha\beta}^{HS}(r_{ij})\right.\nonumber \\
&&
\left.
-\sum_{\alpha}\sum_{i}^{N_{\alpha}}\psi_{\alpha}(r_{i})\right],
\label{2.6}
\end{eqnarray}
where
\begin{equation}
\bar \nu_{\alpha}=\nu_{\alpha}+\frac{1}{2}\sum_{{\mathbf
k}}\tilde\Phi_{\alpha\alpha}^{SR}(k)+\frac{q_{\alpha}^{2}}{2}\sum_{{\mathbf
k}}\tilde\Phi^{C}(k),
\label{2.7}
\end{equation}
\begin{equation}
\psi_{\alpha}(r_{i})={\rm i}\omega(r_{i})+{\rm
i}q_{\alpha}\gamma(r_{i}).
\label{2.8}
\end{equation}

\paragraph{Mean-field approximation.}
 The MF approximation of  functional (\ref{2.5}) is defined by
\begin{equation}
\Xi_{MF}[\nu_{\alpha}]=\exp(-{\cal H}[\nu_{\alpha},\bar\rho, \bar Q, \bar\omega, \bar\gamma]),
\label{2.9}
\end{equation}
where $\bar\rho$, $\bar Q$, $\bar\omega$ and $\bar\gamma$
are the solutions of the saddle point equations:
\begin{eqnarray}
\bar\rho &=&\langle N[\bar \nu_{\alpha};-{\rm i}\bar\omega,-{\rm
i}q_{\alpha}\bar\gamma]\rangle_{HS}, \quad \bar Q=0, \nonumber \\
\bar\omega& =& -{\rm{i}}\bar\rho \tilde{\phi}_{NN}(0), \quad
\bar\gamma=-{\rm{i}} \bar\rho \tilde{\phi}_{NQ}(0).
\label{2.10}
\end{eqnarray}
Substituting (\ref{2.10}) in (\ref{2.9}) we obtain
\[
\Xi_{MF}[\nu_{\alpha}]=\exp\Big[\frac{\beta}{2}\bar \rho^{2}\tilde{\phi}_{NN}(0)\Big]\
\ \Xi_{HS}[\bar \nu_{\alpha};-{\rm i}\bar\omega,-{\rm
i}q_{\alpha}\bar\gamma].
\]
The Legendre transform of $\ln \Xi_{MF}$
gives the free energy in the MF approximation
\begin{eqnarray}
\beta f_{MF}&=&\frac{\beta \mathcal{F}_{MF}}{V}=\beta
f_{HS}(\rho_{\alpha})-\frac{\beta}{2V}\sum_{\alpha}\rho_{\alpha}\sum_{{\mathbf
k}}\tilde\phi_{\alpha\alpha}^{SR}(k)
\nonumber \\
&&-\frac{\beta}{2V}\sum_{\alpha}q_{\alpha}^{2}\rho_{\alpha}\sum_{{\mathbf k}}
\tilde\phi^{C}(k)+\frac{\beta}{2}\rho^{2}\tilde\phi_{NN}(0).
\label{2.10a}
\end{eqnarray}

It is worth noting that for $\phi_{\alpha\beta}^{SR}(r)=0$ we arrive at the free energy of SPM in a zero-loop approximation \cite{caillol}.

\paragraph{Beyond the MF approximation.}
First we present CVs $\rho_{{\mathbf k}}$ and  $Q_{{\mathbf k}}$
($\omega_{{\mathbf k}}$ and $\gamma_{{\mathbf
k}}$) as
\begin{eqnarray*}
\rho_{{\mathbf k}}=\bar\rho \delta_{{\mathbf
k}}+\delta\rho_{{\mathbf k}}, \qquad Q_{{\mathbf k}}=\bar Q
\delta_{{\mathbf k}}+\delta Q_{{\mathbf k}}, \\
\omega_{{\mathbf k}}=\bar\omega \delta_{{\mathbf
k}}+\delta\omega_{{\mathbf k}}, \qquad \gamma_{{\mathbf
k}}=\bar\gamma \delta_{{\mathbf k}}+\delta\gamma_{{\mathbf k}},
\end{eqnarray*}
where the quantaties with a bar are given by (\ref{2.10}) and $\delta_{{\mathbf k}}$ is the Kronecker symbol.

Then we write  $\ln\Xi_{HS}[\bar\nu_{\alpha};-{\rm i}\omega,-{\rm i}q_{\alpha}\gamma]$ (see (\ref{2.6})-(\ref{2.8})) in the form of the cumulant expansion
\begin{eqnarray}
\ln\Xi_{HS}[\ldots]&=&\sum_{n\geq 0}\frac{(-{\rm
i})^{n}}{n!}\sum_{i_{n}\geq 0}
\sum_{{\mathbf{k}}_{1},\ldots,{\mathbf{k}}_{n}}
{\mathfrak{M}}_{n}^{(i_{n})}(k_{1},\ldots,k_{n})\delta\gamma_{{\bf{k}}_{1}}\ldots\delta\gamma_{{\bf{k}}_{i_{n}}}\nonumber \\
&&
\delta\omega_{{\bf{k}}_{i_{n+1}}}\ldots\delta\omega_{{\bf{k}}_{n}}\delta_{{\bf{k}}_{1}+\ldots
+{\bf{k}}_{n}},
\label{2.11}
\end{eqnarray}
where ${\mathfrak{M}}_{n}^{(i_{n})}(k_{1},\ldots,k_{n})$ is the $n$th cumulant  defined by
\begin{equation}
{\mathfrak{M}}_{n}^{(i_{n})}(k_{1},\ldots,k_{n})=\frac{\partial^{n}\ln
\Xi_{HS}[\ldots]}{\partial
\delta\gamma_{{\bf{k}}_{1}}\ldots\partial\delta\gamma_{{\bf{k}}_{i_{n}}}
\partial\delta\omega_{{\bf{k}}_{i_{n+1}}}\ldots\partial\delta\omega_{{\bf{k}}_{n}}}\vert_{\delta\gamma_{{\mathbf{k}}} =0,\delta\omega_{{\mathbf{k}}}=0}.
\label{2.12}
\end{equation}
The expressions for the cumulants (for $n\leq 4$) are given in Appendix~A.

We can integrate in (\ref{2.5}) over $\delta\omega_{{\bf{k}}}$ and $\delta\gamma_{{\bf{k}}}$ and obtain for $\Xi[\nu_{\alpha}]$ (see \cite{patsahan:04:1})
\begin{eqnarray}
\Xi[\nu_{\alpha}]&=&\Xi_{MF}[\bar\nu_{\alpha}] \Xi' \int(\mathrm{d}\delta\rho)(\mathrm{d}\delta
Q)\exp\Big\{-\frac{1}{2!}\sum_{\bf
k}\left[L_{NN}(\bar\nu_{\alpha};k)\delta\rho_{\bf k}\delta\rho_{-\bf
k}\right. \nonumber \\
 &&\left.+2{L_{NQ}}(\bar\nu_{\alpha};k)\delta\rho_{\bf k}\delta Q_{-\bf k}+L_{QQ}(\bar\nu_{\alpha};k)\delta Q_{\bf k}\delta Q_{-\bf k}\right]
\nonumber \\
&&+\sum_{n\geq 3}\sum_{i_{n}\geq
0}\mathcal{H}_{n}^{(i_{n})}(\bar\nu_{\alpha};\delta\rho,\delta Q)\Big\}.
\label{dA.14}
\end{eqnarray}
It is worth noting that the above expression is of the same form as that obtained   within the framework of the mesoscopic field theory \cite{ciach:00:0,ciach:05:0}. The main difference is that $\Xi[\nu_{\alpha}]$ in (\ref{dA.14}) is a function of full chemical potentials, rather than  just a function of their mean field parts.

\subsection{ Gaussian approximation. A two-component primitive model (PM)}

In the Gaussian approximation, which corresponds to taking into account in (\ref{dA.14}) only terms with $n\leq 2$ ($\mathcal{H}_{n}^{(i_{n})}\equiv 0$), we have $L_{AB}(\bar\nu_{\alpha};k)=C_{AB}(\bar\nu_{\alpha};k)$ ($A,B=N,Q$), where
\begin{eqnarray}
C_{NN}&=&\tilde \Phi_{NN}(k)+1/{\mathfrak{M}}_{2}^{(0)}(\bar\nu_{\alpha};k),\quad
C_{QQ}=\tilde \Phi_{QQ}(k)+1/{\mathfrak{M}}_{2}^{(2)}(\bar\nu_{\alpha}), \nonumber \\
C_{NQ}&=&\tilde \Phi_{NQ}(k).
\label{dA.15}
\end{eqnarray}
$C_{NN}(k)$,  $C_{QQ}(k)$ and $C_{NQ}(k)$ are the Fourier transforms of the  density-density, charge-charge and charge-density direct correlation functions, respectively. It is worth noting that
${\mathfrak{M}}_{2}^{(2)}(\bar\nu_{\alpha})$ does not depend on ${\bf k}$ (see Appendix~A).
After integrating in (\ref{dA.14}) over $\delta\rho_{\bf k}$ and $\delta Q_{\bf k}$ taking into account (\ref{dA.15}) we arrive at GPF  in the Gaussian approximation.

Let us consider a two-component  simple primitive model (SPM) \cite{caillol} consisting of
charged hard spheres  of the same diameter ($\sigma_{11}=\sigma_{22}=\sigma$) with $\tilde \Phi_{\alpha\beta}^{SR}(k)=0$  which differ by their respective charges ($z\neq 1$). We  have for SPM (see (\ref{2.3}))
\begin{equation}
\tilde \Phi_{NN}(k)=0, \quad \tilde \Phi_{NQ}(k)=0, \quad \tilde \Phi_{QQ}(k)=\tilde{\Phi}^{C}(k).
\label{RPM}
\end{equation}

It should be noted here that the Hamiltonian of SPM, as for RPM \cite{patsahan:04:1}, does not include a direct pair interaction of number density fluctuations.
Integration over $\delta\rho_{\mathbf k}$  and $\delta\omega_{\mathbf k}$ in (\ref{2.5}) is trivial in this case and  leads to the KSSHE action obtained in \cite{caillol} by performing the Hubbard-Stratonovich transformation. Starting from this expression  the free energy of SPM in a two-loop order approximation was derived by J.-M. Caillol  \cite{caillol}.  As was shown \cite{caillol_1},  the $z$ dependence of  critical temperature calculated within the framework of this approximation disagrees with the results obtained by MC simulations. Below we propose an alternate way of taking into account the fluctuation effects near the GL critical point.

 Expressions  (\ref{dA.15}), under conditions  (\ref{RPM}), are reduced to
\begin{equation}
C_{NN}=1/{\mathfrak{M}}_{2}^{(0)}(\bar\nu_{\alpha};k),\quad
C_{QQ}=\tilde \Phi^{C}(k)+1/{\mathfrak{M}}_{2}^{(2)}(\bar\nu_{\alpha}),\quad
C_{NQ}\equiv 0
\label{dA.16}
\end{equation}
and the logarithm of GPF  in the Gaussian approximation  is as follows:
\begin{equation}
\ln\Xi_{G}(\nu_{\alpha})=\ln \Xi_{\mathrm{HS}}({\bar \nu_{\alpha}})-\frac{1}{2}\sum_{\mathbf{k}}\ln\left[1+\tilde{\Phi}^{C}(k)
{\mathfrak{M}}_{2}^{(2)}({\bar \nu_{\alpha}})\right].
\label{da2.10}
\end{equation}
After the Legendre transform of $\ln\Xi_{G}(\nu_{\alpha})$ we obtain the well-known expression (see e.g. \cite{caillol}) for the free energy in RPA
\begin{equation}
\beta f_{RPA}=\beta f_{MF}+\frac{1}{2V}\sum_{\mathbf k}\ln(1+\kappa_{D}^{2}\phi^{C}(k)),
\label{da2.10a}
\end{equation}
where $\beta f_{MF}$ has the form (\ref{2.10a}) under  conditions (\ref{RPM}),
$\kappa_{D}^{2}=4 \pi \rho\beta q_{0}^{2}z$ is the squared Debye number. It is worth noting that
a use of  momentum cutoff $\vert{\mathbf k}_{\Lambda}\vert=2\pi/a $ in (\ref{da2.10a}) leads to the same expression for the $\beta f_{DH}$  as in  \cite{netz_orland}.
For point charge particles, (\ref{da2.10a}) yields the free energy in the DHLL approximation $\beta f_{DHLL}=\beta f_{id}-\frac{\kappa_{D}^{3}}{12\pi}$.  Using the optimized regularization of the Coulomb potential inside  the hard core \cite{anderson_chandler} we arrive at the free energy in ORPA (or MSA).

As is seen, $\beta f_{RPA}$ does not explicitely depend on the charge asymmetry
factor $z$. The same is true for MSA and the DH theories.
The detailed analysis of the GL phase equilibrium using (\ref{da2.10a})  and  different regularizations of the Coulomb potential inside the hard core was fulfilled  in \cite{caillol_1}.
As was shown \cite{caillol,netz_orland}, the $z$-dependent free energy can be found only in the higher order approximations.

Below we study the GL phase diagram of SPMs when fluctuation effects of the order
higher than the second order are  taken into account.

\section{Gas-liquid critical point of the primitive models (PMs) of ionic fluids}

Let us consider equation (\ref{da2.10}). Introducing  $\bar\nu_{N}$ and $\bar\nu_{Q}$
\begin{equation}
\bar\nu_{N}=\frac{z}{1+z}\bar\nu_{1}+\frac{1}{1+z}\bar\nu_{2}, \qquad
\bar\nu_{Q}=\frac{1}{q_{0}(1+z)}(\bar\nu_{1}-\bar\nu_{2})
\label{2.8a}
\end{equation}
we can present  (\ref{da2.10}) as follows:
\begin{equation}
\ln\Xi_{G}(\nu_{\alpha})=\ln \Xi_{\mathrm{HS}}({\bar \nu_{N}},{\bar \nu_{Q}})-\frac{1}{2}\sum_{\mathbf{k}}\ln\left[1+\tilde{\Phi}^{C}(k)
{\mathfrak{M}}_{2}^{(2)}({\bar \nu_{N}},{\bar \nu_{Q}})\right],
\label{da2.11}
\end{equation}
where new chemical potentials  $\bar\nu_{N}$ and $\bar\nu_{Q}$ (see (\ref{2.8a})) are conjugate to the total number density and charge density, respectively. Since  near the GL critical point the  fluctuations of the number density  play  a crucial role, $\bar\nu_{N}$  is of special interest in this study.

We present $\bar\nu_{N}$ and $\bar\nu_{Q}$  as
\[
\bar\nu_{N}=\nu_{N}^{0}+\lambda\Delta\nu_{N}, \qquad
\bar\nu_{Q}=\nu_{Q}^{0}+\lambda\Delta\nu_{Q}, \qquad
\]
with $\nu_{N}^{0}$ and $\nu_{Q}^{0}$ being the MF values of $\bar\nu_{N}$ and $\bar\nu_{Q}$, respectively and
$\Delta\nu_{N}$ and $\Delta\nu_{Q}$  being the solutions of the equations
\begin{eqnarray}
\frac{\partial\ln \Xi_{G}(\nu_{N},\nu_{Q})}{\partial\Delta\nu_{N}}&=&\lambda\langle N\rangle,
\label{a3.17}
\\
\frac{\partial\ln \Xi_{G}(\nu_{N},\nu_{Q})}{\partial\Delta\nu_{Q}}&=&0.
\label{b3.17}
\end{eqnarray}
Expanding  (\ref{da2.11}) in powers of  $\Delta\nu_{N}$ and $\Delta\nu_{Q}$ we obtain
\begin{equation}
\ln\Xi_{G}(\nu_{N},\nu_{Q})=\sum_{n\geq 0}\sum_{i_{n}\geq 0}^{n}C_{n}^{i_{n}} \frac{{\mathcal
M}_{n}^{(i_{n})}(\nu_{N}^{0},\nu_{Q}^{0})}{n!}\Delta\nu_{Q}^{i_{n}}\Delta\nu_{N}^{n-i_{n}},
\label{a3.16}
\end{equation}
where
\[
{\cal{M}}_{n}^{(i_{n})}(\nu_{N}^{0},\nu_{Q}^{0})=\frac{\partial^{n}\ln\Xi_{G}(\nu_{N},\nu_{Q})}{\partial\Delta\nu_{Q}^{i_{n}}\partial\Delta\nu_{N}^{n-i_{n}}}\mid_{\Delta\nu_{N}=0,\Delta\nu_{Q}=0}.
\]
The expressions for the coefficients ${\mathcal M}_{n}^{(i_{n})}$ are given in Appendix~B.

We solve  equations (\ref{a3.17})-(\ref{b3.17}) approximately taking into account (\ref{a3.16}) and keeping  terms of a certain order in parameter $\lambda$. The  procedure is as follows. First, we calculate $\Delta\nu_{Q}$ from  (\ref{b3.17}) in the approximation which corresponds to a certain order of $\lambda$ e.g., order $s$. Then, we substitute  this $\Delta\nu_{Q}$ into equation (\ref{a3.17}) in order to find  $\Delta\nu_{N}$  in the approximation corresponding to $\lambda^{s+1}$. In  (\ref{a3.17}) we take into account only the linear terms with respect to $\Delta\nu_{N}$ keeping  terms with {\it all powers of}  $\Delta\nu_{Q}$ within a given approximation in terms of $\lambda$.

 As is readily seen, the first nontrivial solution for $\Delta\nu_{N}$ is obtained  in the  approximation of the first order of $\lambda$. It is the result of substitution   in (\ref{a3.17}) of the solution  $\Delta\nu_{Q}=0$.   As a result,  we have
\begin{equation}
\Delta\nu_{N}=-\frac{{\mathfrak{M}}_{3}^{(2)}}{2{\mathfrak{M}}_{2}^{(0)}}\sum_{\mathbf k}\tilde g(k),
\label{RPA_a}
 \end{equation}
where
\begin{equation}
\tilde g(k)=-\frac{\tilde\Phi^{C}(k)}{1+\tilde\Phi^{C}(k)
{\mathfrak{M}}_{2}^{(2)}}=-\frac{1}{V}\frac{\beta \tilde\phi^{C}(k)}{1+\kappa_{D}^{2}\tilde \phi^{C}(k)}.
\label{da3.2}
\end{equation}
  (\ref{RPA_a})-(\ref{da3.2}) can be rewritten as (see  Appendix~C)
\begin{equation}
\Delta\nu_{N}=\frac{1}{2N}\sum_{\mathbf k}\frac{\kappa_{D}^{2}\tilde \phi^{C}(k)}{1+\kappa_{D}^{2}\tilde \phi^{C}(k)}
\label{RPA_b}
\end{equation}
which corresponds to RPA.  As was mentioned above, $\Delta\nu_{N}$ given by (\ref{RPA_b}) does not depend on charge asymmetry factor z. In order to obtain the $z$-dependent expression for the chemical potential related to the number density fluctuations  we should consider the next approximation in $\lambda$ for $\Delta\nu_{N}$.

In order to obtain  $\Delta\nu_{N}$ in the approximation corresponding to $\lambda^{2}$ we substitute  in (\ref{a3.17}) the solution $\Delta\nu_{Q}$ as follows (see  Appendix~C):
\begin{equation}
\Delta\nu_{Q}=-\frac{{\mathfrak{M}}_{3}^{(3)}}{2{\mathfrak{M}}_{2}^{(2)}}\sum_{\mathbf k}\tilde g(k)=-\frac{(1-z)}{2}\sum_{\mathbf k}q_{0}\tilde g(k),
\label{RPA_b1}
\end{equation}
 which is found from (\ref{b3.17})  in the first approximation of $\lambda$. Taking into account only a  linear term with respect to $\Delta\nu_{N}$  we get
\begin{eqnarray}
\Delta\nu_{N}&=&-\frac{1}{{\mathfrak{M}}_{2}^{(0)}}\left[\frac{1}{2}\sum_{\mathbf k}\tilde g(k){\mathfrak{M}}_{3}^{(2)}+\frac{1}{2}\sum_{\mathbf k}\tilde g(k){\mathfrak{M}}_{4}^{(3)}\Delta\nu_{Q}
\right. \nonumber \\
&&\left.
+\frac{1}{2}{\mathfrak{M}}_{3}^{(2)}\Delta\nu_{Q}^{2}+\frac{1}{3!}{\mathfrak{M}}_{4}^{(3)}
\Delta\nu_{Q}^{3}\right].
\label{delta_n}
\end{eqnarray}

Let us consider (\ref{delta_n}) in detail. The correlation effects of the order higher than the second order enter the equation through the cumulants ${\mathfrak{M}}_{n}^{(i_{n})}$  for $n\geq 3$ and $i_{n}\neq 0$. The appearence of these cumulants reflects the fact that the terms proportional to  $\omega\gamma^{2}$, $\gamma^{3}$ and  $\omega\gamma^{3}$ are taking into account  in the cumulant expansion (\ref{2.11}) ($n\leq 4$). Recall that $\omega_{{\mathbf k}}$ and $\gamma_{{\mathbf k}}$ are conjugate to the CVs $\rho_{{\mathbf k}}$ and $Q_{{\mathbf k}}$ describing the total number density and charge density fluctuations, respectively. This means that in order to determine $\Delta\nu_{N}$ we take into account in (\ref{2.11}), besides the terms of the second order,  the contribution corresponding to the pure charge density fluctuations for $n=3$ (${\mathfrak{M}}_{3}^{(3)}$) and the contributions corresponding to the correlations between  charge density and total number density fluctuations  for $n\leq 4$ (${\mathfrak{M}}_{3}^{(2)}$ and ${\mathfrak{M}}_{4}^{(3)}$) which are linear in  $\omega_{{\mathbf k}}$.   Therefore, the analysis of (\ref{delta_n}) establishes a link between the approximations considered above (in terms of $\lambda$) and the approximations formulated in terms of CVs.

Another important issue to be discussed is the limitting case $z=1$ that cooresponds to RPM.
For  $z=1$  only the first term  survives in (\ref{delta_n}). Furthermore, in this case the conditions  ${\mathfrak{M}}_{n}^{(3)}\equiv 0$ and $\Delta\nu_{Q}=0$ hold  simultaneously  (they are equivalent for the RPM) due to the model symmetry (see formulas in appendixes~A and ~C). As a result,  we arrive at the  expression for the chemical potential of RPM in  RPA \cite{caillol_1,patsahan_rpm}.   On the other hand, when we set solely $\Delta\nu_{Q}=0$ in  (\ref{delta_n}) we obtain (\ref{RPA_a}) which also corresponds to RPA and is valid for  $z\neq 1$. The latter reflects the fact that the correlations between  fluctuations of the charge density and the total number density are not taken into account at the RPA level.  Therefore, for $\Delta\nu_{Q}=0$  all the terms, except the first one, in (\ref{delta_n}) become equal to zero and we get the expressions for the chemical potential conjugate to the total number density in the same approximation for the both models. The equation (\ref{delta_n}) is obtained in the approximation when only the liner terms with respect to $\omega$ in the cumulant expansion (\ref{2.11}) are taken into account. However, for  $z=1$ these contributions are equal to zero. This means that the approximation considered in this paper  for SPM does not have an analogy for RPM. In order to include the fluctuations in the simplest model (RPM) the higher order terms  should be taken into account. This issue was considered in \cite{patsahan_rpm} where a good agreement for the critical temperature was obtained.

 Based on (\ref{delta_n})  the coexistence  curves for SPM for differnt values of $z$ can be calculated.
It is worth  noting that the regularization of the potential $\phi_{\alpha\beta}^{C}(r)$  inside the hard core is  arbitrary to some extent.  For example, different regularizations for the Coulomb potential were considered in \cite{ciach:00:0,caillol_1}. Within the framework of the Gaussian approximation of GPF the best estimation for the critical temperature is achieved for the optimized regularization \cite{anderson_chandler} that leads to the ORPA (MSA). However, this approximation does not work properly in the higher orders of the perturbation theory   \cite{caillol_1}.
In this study we use the Weeks-Chandler-Andersen  (WCA) regularization \cite{wcha}. As was shown
\cite{cha}, this choice of $\phi^{C}(r)$ for $r<\sigma$ produces rapid convergence
of the series of the perturbation theory for the free energy. For the WCA, the Fourier transform of
$\phi^{C}(r)$ is of the form $\phi^{C}(x)=\sin(x)/x^{3}$ with $x=k\sigma$.
As a result, (\ref{delta_n}) can be written as (see  Appendix~C)
\begin{eqnarray}
\Delta\nu_{N}=\frac{i_{1}}{\pi}\left[ 1+\frac{i_{1}(1-z)^{2}}{2z\pi}\left( 1-\frac{i_{1}(1-z)^{2}}{3z\pi}\right) \right].
\label{delta_rho_new}
\end{eqnarray}
In (\ref{delta_rho_new})
\begin{equation}
i_{1}=\frac{1}{T^{*}}\int_{0}^{\infty}\frac{x^{2}\sin x dx}{x^{3}+{\kappa^{*}}^{2}\sin x}
\label{a3.15}
\end{equation}
with $\kappa^{*}=\kappa_{D}\sigma$ being the reduced Debye number.
Hereafter the standard notations are introduced for the temperature and the density:    $T^{*}=\frac{k_{B}T\sigma}{zq_{0}^{2}}$, $\rho^{*}=\rho\sigma^{3}$.

Finally, we get the expression for  the full chemical potential $\nu_{N}$ conjugate to the total number density
\begin{eqnarray}
\nu_{N}-3\ln\Lambda/\sigma&=&\ln\rho^{*}+\frac{\eta(8-9\eta+3\eta^{3})}{(1-\eta)^{3}}+\frac{z}{1+z}\ln z-\ln(1+z)  \nonumber \\
&&-\frac{1}{2T^{*}}+\Delta\nu_{N},
\label{nu_full}
\end{eqnarray}
where  $\Delta\nu_{N}$ is given in (\ref{delta_rho_new})-(\ref{a3.15}) and $\eta=\pi\rho^{*}/6$. In (\ref{nu_full}) the Carnahan-Starling  approximation for the hard sphere system is used.

 Fig.~3 shows the coexistence curves for $z=2-4$ calculated  based on  the isotherms of  chemical potential  (\ref{nu_full}) supplemented with the Maxwell construction.
The estimations for the critical point are as follows:
\begin{eqnarray*}
z&=&2: \qquad T_{c}^{*}=0.12310, \qquad \rho_{c}^{*}=0.00946 \\
z&=&3: \qquad T_{c}^{*}=0.11313, \qquad \rho_{c}^{*}=0.02740\\
z&=&4: \qquad T_{c}^{*}=0.10030, \qquad \rho_{c}^{*}=0.04501.
\end{eqnarray*}
 \begin{figure}
\centering
\includegraphics[height=9cm]{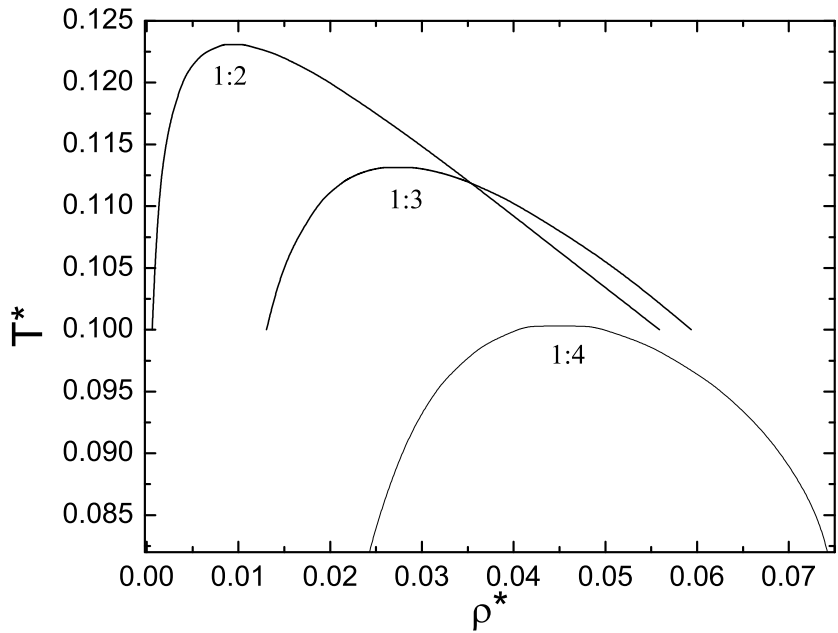}
\caption{Coexistence curves of the ($1:z$) charge-asymmetric ionic model.}
\label{fig2}
\end{figure}
These results demonstrate the qualitative agreement with the MC data: the critical temperature decreases when $z$ increases and the critical density increases rapidly with $z$. Moreover, a comparison of the coexistence curve forms for $z=2$ and $z=3$ (results for the coexistence curve for $z=4$ are not available by now) with the DHBjCINC theory indicates their similarity. It is also worth noting that the above data for the critical temperature lie about in the same region as those obtained in \cite{caillol_1} within the framework of the two-loop expansion.

\section{Conclusions}
In this paper we have studied the GL phase behaviour of a charge asymmetric primitive ionic model.
For this purpose we have derived the exact expression for the functional of GPF of a two-component asymmetric ionic model which includes both the short- and long-range interactions among charged hard spheres.  We have shown that  the well-known approximations for the free energy, in particular DHLL and ORPA, can be reproduced within the framework of this approach. On the other hand, the GPF functional can be reduced to the form found in the KSSHE theory \cite{caillol}. This means that the field-theoretical analysis of the expression for GPF given by (\ref{2.5})-(\ref{2.5a})   has to lead  in the two-loop approximation to the same $z$-dependence  for the critical temperatures as in \cite{caillol_1} which does not agree with the MC simulations. Here, we have proposed an alternative method for the study of the GL phase equilibria in SPM. It consists in the calculation of the chemical potential $\nu_{N}$ conjugate to the total number density by means of successive approximations.

In conclusion, we have obtained the expression for the chemical potential $\nu_{N}$ in which the effects of indirect correlations between the number density fluctuations are taken into consideration via a charge subsystem. Based on this expression supplemented with the Maxwell constraction  the coexistence curves for $z=2-4$ have been  calculated. The $z$-dependences obtained for both the critical temperature and the critical density qualitatively agree  with MC simulations. The results demonstrate that the terms responsible for the charged clustering need not  be included into the Hamiltonian explicitely in order to describe, at least  qualitatively, the GL phase diagram of SPM. Instead, the interaction between charge and density fluctuations should be properly taken into account. To achieve a quantitative agreement, the higher order terms  should be taken into consideration in the CV action. This task will be considered elsewhere.

\section*{Acknowledgement}
 The authors would like to thank A. Ciach and J.-M. Caillol for useful discussions. This work was partly supported by the cooperation project  between the CNRS and the NASU entitled "Effects of asymmetry on phase diagrams and dynamics of fluid mixtures".

\section*{Appendix~A:  Recurrence formulas for the cumulants in the Fourier space representation}
\begin{eqnarray}
{\mathfrak{M}}_{n}^{(0)}(k_{1}, k_{2},\ldots, k_{n})&=& {\widetilde G}_{n}(k_{1},k_{2},\ldots, k_{n})
\label{3.14a} \\
{\mathfrak{M}}_{n}^{(1)}(k_{1}, k_{2},\ldots, k_{n})&=&0 \label{3.14b} \\
{\mathfrak{M}}_{n}^{(2)}(k_{1}, k_{2},\ldots,k_{n})&=&
\sum_{\alpha}q_{\alpha}^{2}c_{\alpha}{\widetilde G}_{n-1}(k_{1}, k_{2},\ldots,\vert{\mathbf k}_{n-1}+{\mathbf k}_{n}\vert) \label{3.14c} \\
{\mathfrak{M}}_{n}^{(3)}(k_{1}, k_{2},\ldots, k_{n})&=&
\sum_{\alpha}q_{\alpha}^{3}c_{\alpha}{\widetilde G}_{n-2}(k_{1},k_{2},\ldots,\vert{\mathbf k}_{n-2}+{\mathbf k}_{n-1}+{\mathbf k}_{n}\vert)
\label{3.14d}
\end{eqnarray}
where ${\widetilde G}_{n}(k_{1}, k_{2},\ldots, k _{n})$ is the Fourier transform of the $n$-particle  truncated correlation function \cite{stell} of a one-component hard sphere system.

\section*{Appendix~B:  Explicit expressions for ${\cal{M}}_{n}^{(i_{n})}$}
\begin{eqnarray*}
{\cal{M}}_{1}^{(0)}&=&\lambda{\mathfrak{M}}_{1}^{(0)}+\frac{\lambda^{3}}{2}{\mathfrak{M}}_{3}^{(2)}\sum_{\mathbf{k}}\tilde{g}(k),
\nonumber
\\
{\cal{M}}_{1}^{(1)}&=&\frac{\lambda^{3}}{2}{\mathfrak{M}}_{3}^{(3)}\sum_{\mathbf{k}}\tilde{g}(k),
\\
{\cal{M}}_{2}^{(0)}&=&\lambda^{2}{\mathfrak{M}}_{2}^{(0)}+\frac{\lambda^{4}}{2}{\mathfrak{M}}_{4}^{(2)}
\sum_{\mathbf{k}}\tilde{g}(k)+\frac{\lambda^{6}}{2}\Big[{\mathfrak{M}}_{3}^{(2)}\Big]^{2}
\sum_{\mathbf{k}}\tilde{g}^{2}(k),
\\
{\cal{M}}_{2}^{(1)}&=&\frac{\lambda^{4}}{2}{\mathfrak{M}}_{4}^{(3)}
\sum_{\mathbf{k}}\tilde{g}(k)+\frac{\lambda^{6}}{2}{\mathfrak{M}}_{3}^{(2)}
{\mathfrak{M}}_{3}^{(3)}\sum_{\mathbf{k}}\tilde{g}^{2}(k),
\\
{\cal{M}}_{2}^{(2)}&=&\lambda^{2}{\mathfrak{M}}_{2}^{(2)}+\frac{\lambda^{4}}{2}{\mathfrak{M}}_{4}^{(4)}
\sum_{\mathbf{k}}\tilde{g}(k)+\frac{\lambda^{6}}{2}\Big[{\mathfrak{M}}_{3}^{(3)}\Big]^{2}
\sum_{\mathbf{k}}\tilde{g}^{2}(k),
\end{eqnarray*}

\begin{eqnarray*}
{\cal{M}}_{3}^{(0)}&=&\lambda^{3}{\mathfrak{M}}_{3}^{(0)}+\frac{\lambda^{5}}{2}{\mathfrak{M}}_{5}^{(2)}\sum_{\mathbf{k}}\tilde{g}(k)+\frac{3\lambda^{7}}{2}{\mathfrak{M}}_{3}^{(2)}
{\mathfrak{M}}_{4}^{(2)}\sum_{\mathbf{k}}\tilde{g}^{2}(k)\nonumber
\\
&&
+\lambda^{9}\Big[{\mathfrak{M}}_{3}^{(2)}(\nu_{N}^{0},\nu_{Q}^{0})\Big]^{3}\sum_{\mathbf{k}}\tilde{g}^{3}(k),
\\
{\cal{M}}_{3}^{(1)}&=&\frac{\lambda^{5}}{2}{\mathfrak{M}}_{5}^{(2)}\sum_{\mathbf{k}}\tilde{g}(k)+\frac{\lambda^{7}}{2}\left[{\mathfrak{M}}_{3}^{(3)}
{\mathfrak{M}}_{4}^{(2)}+{\mathfrak{M}}_{3}^{(2)}
{\mathfrak{M}}_{4}^{(3)}\right] \sum_{\mathbf{k}}\tilde{g}^{2}(k)
\nonumber
\\
&&
+\lambda^{9}\Big[{\mathfrak{M}}_{3}^{(2)}\Big]^{2}{\mathfrak{M}}_{3}^{(3)}\sum_{\mathbf{k}}\tilde{g}^{3}(k),
\\
{\cal{M}}_{3}^{(2)}&=&\lambda^{3}{\mathfrak{M}}_{3}^{(2)}+\frac{\lambda^{5}}{2}{\mathfrak{M}}_{5}^{(4)}\sum_{\mathbf{k}}\tilde{g}(k)+\frac{\lambda^{7}}{2}\left[{\mathfrak{M}}_{3}^{(2)}
{\mathfrak{M}}_{4}^{(4)}+2{\mathfrak{M}}_{3}^{(3)}(\nu_{N}^{0},\nu_{Q}^{0})\right.
\nonumber
\\
&&
\left.
\times{\mathfrak{M}}_{4}^{(3)}(\nu_{N}^{0},\nu_{Q}^{0})\right]\sum_{\mathbf{k}}\tilde{g}^{2}(k)+\lambda^{9}\Big[{\mathfrak{M}}_{3}^{(3)}(\nu_{N}^{0},\nu_{Q}^{0})\Big]^{2}{\mathfrak{M}}_{3}^{(2)}(\nu_{N}^{0},\nu_{Q}^{0})\nonumber
\\
&&
\times\sum_{\mathbf{k}}\tilde{g}^{3}(k),
\\
{\cal{M}}_{3}^{(3)}&=&\lambda^{3}{\mathfrak{M}}_{3}^{(3)}+\frac{\lambda^{5}}{2}{\mathfrak{M}}_{5}^{(5)}\sum_{\mathbf{k}}\tilde{g}(k)+\frac{3\lambda^{7}}{2}{\mathfrak{M}}_{3}^{(3)}
{\mathfrak{M}}_{4}^{(4)}\sum_{\mathbf{k}}\tilde{g}^{2}(k)\nonumber
\\
&&
+\lambda^{9}\Big[{\mathfrak{M}}_{3}^{(3)}\Big]^{3}\sum_{\mathbf{k}}\tilde{g}^{3}(k),
\end{eqnarray*}

\begin{eqnarray*}
{\cal{M}}_{4}^{(0)}&=&\lambda^{4}{\mathfrak{M}}_{4}^{(0)}+\frac{\lambda^{6}}{2}{\mathfrak{M}}_{6}^{(2)}
\sum_{\mathbf{k}}\tilde{g}(k)+\frac{\lambda^{8}}{2}\Big(3\Big[{\mathfrak{M}}_{4}^{(2)}\Big]^{2}
+4{\mathfrak{M}}_{3}^{(2)}{\mathfrak{M}}_{5}^{(2)}\Big)\nonumber
\\
&&
\times\sum_{\mathbf{k}}\tilde{g}^{2}(k)+6\lambda^{10}\Big[{\mathfrak{M}}_{3}^{(2)}\Big]^{2}
{\mathfrak{M}}_{4}^{(2)}\sum_{\mathbf{k}}\tilde{g}^{3}(k)+3\lambda^{12}\Big[{\mathfrak{M}}_{3}^{(2)}\Big]^{4}\nonumber
\\
&&
\times\sum_{\mathbf{k}}\tilde{g}^{4}(k),
\\
{\cal{M}}_{4}^{(1)}&=&\frac{\lambda^{6}}{2}{\mathfrak{M}}_{6}^{(3)}
\sum_{\mathbf{k}}\tilde{g}(k)+\frac{\lambda^{8}}{2}\Big(3{\mathfrak{M}}_{4}^{(2)}{\mathfrak{M}}_{4}^{(4)}
+{\mathfrak{M}}_{3}^{(2)}{\mathfrak{M}}_{5}^{(3)}+{\mathfrak{M}}_{3}^{(3)}{\mathfrak{M}}_{5}^{(2)}\Big)\nonumber
\\
&&
\times\sum_{\mathbf{k}}\tilde{g}^{2}(k)+3\lambda^{10}\left({\mathfrak{M}}_{3}^{(2)}{\mathfrak{M}}_{3}^{(3)}{\mathfrak{M}}_{4}^{(4)}+\Big[{\mathfrak{M}}_{3}^{(2)}\Big]^{2}
{\mathfrak{M}}_{4}^{(3)}\right)\sum_{\mathbf{k}}\tilde{g}^{3}(k)
\nonumber
\\
&&
+3\lambda^{12}\Big[{\mathfrak{M}}_{3}^{(2)}\Big]^{3}{\mathfrak{M}}_{3}^{(3)}\sum_{\mathbf{k}}\tilde{g}^{4}(k),
\\
{\cal{M}}_{4}^{(2)}&=&\lambda^{4}{\mathfrak{M}}_{4}^{(2)}+\frac{\lambda^{6}}{2}{\mathfrak{M}}_{6}^{(4)}
\sum_{\mathbf{k}}\tilde{g}(k)+\frac{\lambda^{8}}{2}\Big(2{\mathfrak{M}}_{3}^{(3)}{\mathfrak{M}}_{5}^{(3)}+2{\mathfrak{M}}_{3}^{(2)}{\mathfrak{M}}_{5}^{(3)}
\nonumber
\\
&&
+{\mathfrak{M}}_{4}^{(2)}{\mathfrak{M}}_{4}^{(4)}+2\Big[{\mathfrak{M}}_{4}^{(3)}\Big]^{2}+2{\mathfrak{M}}_{4}^{(2)}\Big[{\mathfrak{M}}_{3}^{(3)}\Big]^{2}\Big)\sum_{\mathbf{k}}\tilde{g}^{2}(k)+2\lambda^{10}\nonumber
\\
&&
\times\Big(2{\mathfrak{M}}_{3}^{(2)}{\mathfrak{M}}_{3}^{(3)}{\mathfrak{M}}_{4}^{(3)}+\Big[{\mathfrak{M}}_{3}^{(2)}\Big]^{2}
{\mathfrak{M}}_{4}^{(4)}\Big)\sum_{\mathbf{k}}\tilde{g}^{3}(k)+3
\lambda^{12}\Big[{\mathfrak{M}}_{3}^{(2)}\Big]^{2}
\nonumber
\\
&&
\times\Big[{\mathfrak{M}}_{3}^{(3)}\Big]^{2}\sum_{\mathbf{k}}\tilde{g}^{4}(k),
\\
{\cal{M}}_{4}^{(3)}&=&\lambda^{4}{\mathfrak{M}}_{4}^{(2)}+\frac{\lambda^{6}}{2}{\mathfrak{M}}_{6}^{(5)}
\sum_{\mathbf{k}}\tilde{g}(k)+\frac{\lambda^{8}}{2}\Big(3{\mathfrak{M}}_{3}^{(3)}{\mathfrak{M}}_{5}^{(4)}
+{\mathfrak{M}}_{3}^{(2)}{\mathfrak{M}}_{5}^{(5)}\nonumber
\\
&&
+3{\mathfrak{M}}_{4}^{(3)}{\mathfrak{M}}_{4}^{(4)}
\Big)\sum_{\mathbf{k}}\tilde{g}^{2}(k)+3\lambda^{10}\nonumber
\Big({\mathfrak{M}}_{3}^{(2)}{\mathfrak{M}}_{3}^{(3)}{\mathfrak{M}}_{4}^{(4)}+\Big[{\mathfrak{M}}_{3}^{(3)}\Big]^{2}{\mathfrak{M}}_{4}^{(3)}\Big)\nonumber
\\
&&
\times\sum_{\mathbf{k}}\tilde{g}^{3}(k)+3\lambda^{12}{\mathfrak{M}}_{3}^{(2)}
{\mathfrak{M}}_{5}^{(5)}\sum_{\mathbf{k}}\tilde{g}^{4}(k).
\end{eqnarray*}

In the above formulas ${\mathfrak{M}}_{n}^{(i_{n})}={\mathfrak{M}}_{n}^{(i_{n})}(\nu_{N}^{0},\nu_{Q}^{0})$.

\section*{Appendix~C: Some explicit relations obtained for a $1:z$ asymmetric PM}
Let us consider the expressions (\ref{3.14a})-(\ref{3.14d}). For $1:z$ asymmetric PM we have
\[
\sum_{\alpha}q_{\alpha}^{2}c_{\alpha}=q_{0}^{2}z, \qquad
\sum_{\alpha}q_{\alpha}^{3}c_{\alpha}=q_{0}^{3}z(1-z).
\]
Using the above relations and (\ref{3.14a})-(\ref{3.14d}) we can obtain the following explicit relations for  SPM
\[
\frac{{\mathfrak{M}}_{3}^{(2)}}{{\mathfrak{M}}_{2}^{(0)}}=q_{0}^{2}z,
\quad
\left( \frac{{\mathfrak{M}}_{3}^{(3)}}{{\mathfrak{M}}_{2}^{(2)}}\right) ^{2}=q_{0}^{2}z\frac{(1-z)^{2}}{z},
\quad
\frac{{\mathfrak{M}}_{4}^{(3)}{\mathfrak{M}}_{3}^{(3)}}{{\mathfrak{M}}_{2}^{(2)}{\mathfrak{M}}_{3}^{(2)}}=q_{0}^{2}z\frac{(1-z)^{2}}{z},
\]
\[
\left( \frac{{\mathfrak{M}}_{3}^{(3)}}{{\mathfrak{M}}_{2}^{(2)}}\right) ^{3}\frac{{\mathfrak{M}}_{4}^{(3)}}{{\mathfrak{M}}_{3}^{(2)}}=\left(q_{0}^{2}z\frac{(1-z)^{2}}{z} \right)^{2}.
\]

\end{document}